\documentclass[aps,prl,floatfix,nofootinbib,twocolumn,superscriptaddress]{revtex4}
\usepackage{amsmath, amsfonts, amsthm, amssymb, graphicx}
\usepackage{epstopdf}

\def\be{\begin{equation}}
\def\ee{\end{equation}}
\def\ba{\begin{eqnarray}}
\def\ea{\end{eqnarray}}
\def\del{\partial}

\begin{document}

\title{General second order scalar-tensor theory, self tuning, and  {\it the Fab Four}}

\author{Christos Charmousis} 
\affiliation{LPT, CNRS UMR 8627, Universit\'e Paris Sud-11, 91405 Orsay Cedex, France.}
\affiliation{LMPT, CNRS UMR 6083, Universit\'e Fran\c{c}ois Rabelais-Tours, 37200, France} 
\author{Edmund J. Copeland} 
\author{Antonio Padilla} 
\author{Paul M. Saffin} 
\affiliation{School of Physics and Astronomy, 
University of Nottingham, Nottingham NG7 2RD, UK}

\date{\today}

\begin{abstract}
Starting from the most general scalar-tensor theory with second order field equations in four dimensions, we establish the unique action that will allow for the existence of a consistent self-tuning mechanism on FLRW backgrounds, and show how it can be understood as a combination of just four base Lagrangians with an intriguing geometric structure dependent on  the Ricci scalar, the Einstein tensor, the double dual of the Riemann tensor and the Gauss-Bonnet combination. Spacetime curvature can be screened from the net cosmological constant at any given moment because we allow the scalar field to break Poincar\'e invariance on the self-tuning vacua, thereby evading the Weinberg no-go theorem. We show how the four arbitrary functions of the scalar field combine in an elegant way opening up the possibility of obtaining non-trivial cosmological solutions. 
\end{abstract}


\maketitle

In a little known paper published in 1974, G.W. Horndeski presented the most general scalar-tensor theory with second order field equations in four dimensions \cite{horny}.  Given the amount of research into modified gravity over the last ten years or so (see \cite{review} for a review), it seems appropriate to revisit Horndeski's work. Scalar tensor models of modified gravity range from Brans-Dicke gravity \cite{bdgravity} to the recent models \cite{covgal,galmodels} inspired by galileon theory \cite{galileon}, the latter being examples of  higher order scalar tensor Lagrangians with second order field equations. Each of these models represent a special case of Horndeski's panoptic theory.

In this letter, we study Horndeski's theory on FLRW backgrounds. In particular we ask whether or not there are subclasses of \cite{horny} giving a viable self-tuning mechanism for solving the (old) cosmological constant problem. In other words, we ask if one can completely screen the spacetime curvature from the net cosmological constant. Naively one might expect this to be impossible on account of Weinberg's no-go theorem \cite{nogo}. However, Weinberg not only assumes  Poincar\'e invariance at the level of the  spacetime curvature but also at the level of the self-adjusting fields. Here we follow a route similar to \cite{bigalileon} and allow our scalar field to break Poincar\'e invariance on the self-tuning vacua, whilst maintaining a flat spacetime geometry.

By demanding that the self-tuning mechanism continues to work through phase transitions that cause the vacuum energy to jump, we are able to impose some powerful restrictions on Horndeski's theory. Based on equivalence principle (EP) considerations, we assume that matter is only minimally coupled to the metric and then pass the model through our self-tuning filter. This reduces it to four base Lagrangians each depending on an arbitrary function of the scalar only. We call these base Lagrangians {\it the Fab Four},
\begin{eqnarray}
\label{eq:john}
{\cal L}_{john} &=& \sqrt{-g} V_{john}(\phi)G^{\mu\nu} \nabla_\mu\phi \nabla_\nu \phi \\
\label{eq:paul}
{\cal L}_{paul} &=&\sqrt{-g}V_{paul}(\phi)   P^{\mu\nu\alpha \beta} \nabla_\mu \phi \nabla_\alpha \phi \nabla_\nu \nabla_\beta \phi \\
\label{eq:george}
{\cal L}_{george} &=&\sqrt{-g}V_{george}(\phi) R\\
\label{eq:ringo}
{\cal L}_{ringo} &=& \sqrt{-g}V_{ringo}(\phi) \hat G
\end{eqnarray}
where $\hat G=R_{\mu\nu \alpha \beta}R^{\mu\nu \alpha \beta}-4R_{\mu\nu}R^{\mu\nu}+R^2$ is the Gauss-Bonnet combination, $\varepsilon_{\mu\nu\alpha \beta}$ is the Levi-Civita tensor and 
$P^{\mu\nu\alpha \beta} =-\frac{1}{4}\varepsilon^{\mu\nu \lambda \sigma } \;R_{\lambda \sigma \gamma \delta  } \; \varepsilon^{\alpha\beta \gamma \delta}$
is the double dual of the Riemann tensor \cite{mtw}. 

Our results prove that any self tuning scalar-tensor theory (satisfying EP) must be built from the Fab Four.  The weakest of the four is Ringo since this cannot give rise to self-tuning without ``a little help from [its] friends", John and  Paul. When this is the case, Ringo does have a non-trivial  effect on the cosmological  dynamics  but does not spoil self-tuning. George also has difficulties in going solo: when $V_{george}=$ const., we just have GR  and no  self-tuning, whereas when $V_{george}\neq $ const., we have Brans-Dicke gravity with $w=0$, which does self-tune but  is immediately ruled out by solar system constraints.
Thus it is best to consider {\it the Fab Four} as combining to give a single theory, as opposed to four different theories in their own right. In particular we expect that one should always include John and/or Paul for the reasons given above, and because their non-trivial derivative interactions might give rise to Vainshtein effects \cite{vainsh}  that would help in passing solar system tests. Chameleon effects \cite{cham} may also play an important role in this regard.  
\section{Horndeski's theory}\label{appA}
The most  general second-order scalar tensor theory is 
\be 
\label{eq:action}
S=S_{H}[g_{\mu\nu}, \phi]+S_m[g_{\mu\nu}; \Psi_n]
\ee
 where the Horndeski action, $S_{H}= \int d^4 x \sqrt{-g} {\cal L}_{H}$, is obtained from equation  (4.21) of \cite{{horny}},
\begin{widetext}
\ba
\label{eq:hornyLagrangian}
{\cal L}_{H} &=& \delta^{ijk}_{\mu\nu\sigma}\left[\kappa_1\nabla^\mu\nabla_i\phi R_{jk}^{\;\;\;\;\nu\sigma}
           -\frac{4}{3}\kappa_{1,\rho}\nabla^\mu\nabla_i\phi\nabla^\nu\nabla_j\phi\nabla^\sigma\nabla_k\phi        +\kappa_3\nabla_i\phi\nabla^\mu\phi R_{jk}^{\;\;\;\;\nu\sigma}
           -4\kappa_{3,\rho}\nabla_i\phi\nabla^\mu\phi\nabla^\nu\nabla_j\phi\nabla^\sigma\nabla_k\phi \right]\\\nonumber
        &~&+\delta_{\mu\nu}^{ij}\left[(F+2W)R_{ij}^{\;\;\;\;\mu\nu}
           -4F_{,\rho}\nabla^\mu\nabla_i\phi\nabla^\nu\nabla_j\phi+2\kappa_8\nabla_i\phi\nabla^\mu\phi\nabla^\nu\nabla_j\phi\right]
        -3[2(F+2W)_{,\phi}+\rho\kappa_8]\nabla_\mu\nabla^\mu\phi
          +\kappa_9(\phi,\rho),
\ea
\end{widetext}
 with $\rho=\nabla_\mu\phi\nabla^\mu\phi$ and $\delta^{i_1i_2...i_n}_{\mu_1\mu_2...\mu_n}=n!\delta^{[i_1}_{\mu_1}\delta^{i_2}_{\mu_2}...\delta^{i_n]}_{\mu_n}$.  We have  four arbitrary functions of $\phi$ and $\rho$,   $\kappa_i=\kappa_i(\phi,\rho)$  as well as $F=F(\phi, \rho)$, which  is constrained so that
$
F_{,\rho}=\kappa_{1,\phi}-\kappa_3-2\rho\kappa_{3,\rho}.
$
Note that $W=W(\phi)$, which means that it can be absorbed into a redefinition of $F(\phi, \rho)$. The matter part of the action is given by $S_m[g_{\mu\nu}; \Psi_n]$, where we require that the matter fields, $\Psi_n$, are all minimally coupled to the metric $g_{\mu\nu}$.  This follows (without further  loss of generality) from assuming that there is no violation of equivalence principle\footnote{For EP to hold all matter must be minimally coupled to the {\it same} metric, $\tilde g_{\mu\nu}$, and this should only be a function of $g_{\mu\nu}$ and $\phi$. Dependence  on derivatives is not allowed since it would result in the  gravitational coupling to matter being momentum dependent, leading to violations of EP.  Given $\tilde g_{\mu\nu}=\tilde g_{\mu\nu}(g_{\alpha\beta}, \phi)$, we simply compute $g_{\alpha\beta}=g_{\alpha\beta}(\tilde g_{\mu\nu}, \phi)$, and substitute back into the action (\ref{eq:action}), before dropping the tildes. Since this procedure will not generate any additional derivatives in the equations of motion, it  simply serves to  redefine the Horndeski potentials, $\kappa_i(\phi, \rho)$. }. This reasoning  is consistent with the original construction of   Brans-Dicke gravity~\cite{bdgravity}. 

Here we are interested in Horndeski's theory on FLRW backgrounds, for which we have a homogeneous scalar, $\phi=\phi(t)$, and a homogeneous and isotropic metric, 
\be
g_{\mu\nu} dx^\mu dx^\nu =-dt^2+a^2(t)\left[\frac{dr^2}{1-\kappa r^2}+r^2\;d\Omega_{(2)}\right]
\ee
with $\kappa$ being a (positive or negative) constant, specifying the spatial curvature. Plugging this into (\ref{eq:hornyLagrangian}), we obtain an effective Horndeski Lagrangian in the minisuperspace approximation 
\be
\label{eq:effectiveLag}
L_H^\textrm{eff}(a, \dot a, \phi, \dot \phi)=a^3\sum_{n=0}^3\left(X_n-Y_n\frac{\kappa}{a^2}\right)H^n
\ee
where $H=\dot a/a$ is the Hubble parameter,  and we have,
\begin{multline}
X_0 =-\tilde{Q}_{7,\phi}\dot\phi+\kappa_9, \quad X_1 =3(2\kappa_8\dot\phi^3-4F_{,\phi}\dot\phi+Q_7\dot\phi-\tilde{Q}_7),   \\
X_2 =-12(F+F_{,\rho}\dot\phi^2), \qquad  X_3  =8\kappa_{1,\rho}\dot\phi^3,  \\
Y_0=\tilde Q_{1,\phi}\dot\phi+ 12\kappa_3\dot\phi^2-12F, \qquad
  Y_1=\tilde{Q}_1-Q_1\dot\phi  \nonumber
\end{multline}
where 
$
Q_1 = \del \tilde{Q}_1/\del\dot\phi=-12\kappa_1$ and $Q_7 =\del \tilde{Q}_7/\del\dot\phi=6F_{,\phi}-3\dot\phi^2\kappa_8
$.
In a cosmological setting, the matter action contributes a homogeneous and isotropic fluid with energy density $\rho_m$ and pressure $p_m$, satisfying the usual conservation law $\dot \rho_m+3H(\rho_m+p_m)=0$.

The generalized Friedmann equation follows in the standard manner  by computing the Hamiltonian density for the Horndeski Lagrangian, and identifying it with the energy density, $\rho_m$, as follows
\be \label{eq:H}
{\cal H}(a, \dot a, \phi, \dot \phi)=\frac{1}{a^3}\left[\dot a \frac{\del L_H^\textrm{eff}}{\del \dot a}+\dot \phi \frac{\del L_H^\textrm{eff}}{\del \dot \phi}-L_H^\textrm{eff}\right]=-\rho_m
\ee
Since matter only couples directly to the metric, and not the scalar, the scalar equation of motion is given by
\begin{multline} 
{\cal E}(a, \dot a, \ddot a, \phi, \dot \phi, \ddot \phi)= \frac{d}{dt} \left[\frac{\del L_H^\textrm{eff}}{\del \dot  \phi}\right]-\frac{\del L_H^\textrm{eff}}{\del \phi}=0\label{eq:E}
\end{multline}
Note that this equation is always linear in both $\ddot a$ and $\ddot \phi$.
\section{Self-tuning}
Since our ultimate goal is to identify those corners of Horndeski's theory that exhibit self-tuning, we first ask what it means to self-tune, in a relatively model independent way.  Consider our cosmological background in vacuum. The matter sector is expected to contribute a constant vacuum energy density, which we identify with the cosmological constant, $\langle \rho_{m}\rangle_\textrm{vac} =\rho_\Lambda$. In a self tuning scenario, this should not impact on the curvature, so whatever the value of $\rho_\Lambda$,  we have  a Minkowski spacetime\footnote{Different values of $\kappa$ represent different slicings of Minkowksi space: $\kappa=0$ corresponds to Minkowski coordinates; $\kappa<0$ corresponds to Milne coordinates; $\kappa>0$ is not permitted since one cannot foliate Minkowski space with spherical spatial sections.}, with $H^2+\kappa/a^2=0$. This should remain true even when the matter sector goes through a phase-transition, changing the overall value of $\rho_\Lambda$ by a constant amount. This extra requirement will place the strongest constraints on our theory.

In order to proceed we shall take these transitions to be instantaneous, thereby assuming that $\rho_\Lambda$ evolves in a piecewise constant fashion. Now consider a self-tuning solution, $H^2+\kappa/a^2=0$, $\phi=\phi_\Lambda(t)$, satisfying the ``on-shell-in-a'' \footnote{By "on-shell-in-$a$" we mean that we have set $H^2=-\kappa/a^2$.We shall see later that this is indeed a consistent solution to our system.} equations of motion for the metric
\be \label{eq:onshH}
\bar {\cal H}(\phi_\Lambda, \dot \phi_\Lambda, a) = -\rho_\Lambda 
\ee
and the scalar,
\be \label{eq:onshE}
\bar {\cal E}(\phi_\Lambda, \dot \phi_\Lambda, \ddot \phi_\Lambda,  a)= \ddot\phi_{\Lambda} f(\phi_\Lambda,\dot\phi_\Lambda,a)+g(\phi_\Lambda,\dot\phi_\Lambda ,a)=0
\ee
Suppose that a phase transition occurs at some arbitrary time $t=t_*$, so that $ \rho_\Lambda(t_*^-) \neq \rho_{  \Lambda}(t_*^+)$. We require that the scalar field is continuous at the transition,  $\phi_\Lambda(t_*^-)=\phi_{\Lambda}(t_*^+)$, but allow  its derivative to jump, $\dot \phi_\Lambda(t_*^-) \neq \dot \phi_{  \Lambda}(t_*^+)$.
 We first consider equation (\ref{eq:onshH}). This is discontinuous on the right hand side, so it must also be discontinuous on the left, which means that $\bar{\cal H}$ {\it must} have some non-trivial $\dot \phi_\Lambda$ dependence. 
Next consider equation (\ref{eq:onshE}). As $\dot \phi_\Lambda$ is discontinuous, $\ddot \phi_\Lambda$ must run into a delta function at $t=t_*$. This is not supported on the right hand side of equation (\ref{eq:onshE}), and  since $t_*$ can be chosen arbitrarily, we deduce that  $f$ must vanish independently of $g$, so that  (\ref{eq:onshE}) actually splits into two equations
\be
f(\phi_\Lambda,\dot\phi_\Lambda,a)=0, \qquad g(\phi_\Lambda,\dot\phi_\Lambda,a)=0
\ee
Focusing on the former it is clear that if $f$ has nontrivial dependence of $\dot \phi_\Lambda$ then it may be discontinuous at the transition. Since it is constrained to vanish either side of the transition we deduce that $\frac{\del f}{\del \dot \phi_\Lambda}=0$, or equivalently $f=f(\phi_\Lambda, a)$.  Using this simplified form for $f$, we now take derivatives, staying on-shell-in-a, so that we have
\be
\frac{df}{dt}(\phi_\Lambda,\dot\phi_\Lambda,a)=\frac{\del f}{\del \phi_\Lambda}\dot \phi_\Lambda+\frac{\del f}{\del a}\sqrt{-\kappa}=0
\ee
Again, applying similar logic we now conclude that $\frac{\del f}{\del  \phi_\Lambda}=0$ or equivalently $f=f(a)$. An identical line of argument implies that $g=g(a)$. What this tells us is that the on-shell-in-a scalar equation of motion   (\ref{eq:onshE}) has lost all dependence on the scalar field $\phi_\Lambda$ and its derivatives. $\phi_\Lambda(t)$ is fixed by the gravity equation (\ref{eq:onshH}), and  must necessarily retain some non-trivial  time dependence even away from transitions in order to evade the clutches of Weinberg's theorem. More generally, in order to cope with transitions the on-shell-in-a gravity equation (\ref{eq:onshH}) must depend on $\dot \phi_\Lambda$. The scalar equation (\ref{eq:E}) should vanish identically on a flat spacetime, and  must therefore have the schematic form.
\be
{\cal E}= \sum_{n\geq 1}\left[ A_n+\tilde A_n \frac{d}{dt}\right] \Delta_n 
\ee
where $A_n=\ddot \phi  \alpha(\phi, \dot \phi, a)+\beta(\phi, \dot \phi, a),~\tilde A_n=\ddot \phi  \tilde\alpha(\phi, \dot \phi, a)+\tilde\beta(\phi, \dot \phi, a)$ and we define
\be
\Delta_n=H^n-\left(\frac{\sqrt{-\kappa}}{a}\right)^n
\ee
which vanishes on-shell-in-a for $n>0$. Since  the scalar equation  (\ref{eq:E}) ultimately forces self-tuning, it should not be trivial. Furthermore, for a remotely viable cosmology it should be dynamical in the sense that we can {\it evolve}  towards $H^2+\kappa/a^2=0$ rather than having it be true at all times. This imposes the condition that at least one of the $\tilde A_n$ should be non-vanishing. Note that the sum does not include $n=0$, which is absolutely crucial in order to force self-tuning

Let us now apply the self-tuning filters to Horndeski's theory. Using equation (\ref{eq:E}) we can  infer the following form of the minisuperspace Lagrangian in a self tuning set-up, 
\be
\label{eq:selftunlag}
L_\textrm{self-tun}^\textrm{eff}=a^3\left[c(a)+\sum_{n=1}^3 Z_n(\phi, \dot \phi, a) \Delta_n \right].
\ee
In order for the on-shell-in-a gravity equation (\ref{eq:onshH}) to retain dependence on $\dot \phi_\Lambda$ we demand that 
$
\sum_{n=1}^3 nZ_{n, \dot \phi} \left(\frac{\sqrt{-\kappa}}{a}\right)^n  \neq 0
$.
By requiring (\ref{eq:effectiveLag}) to take the form (\ref{eq:selftunlag}) up to a total derivative, we find that we must have $\kappa<0$, and that
\ba
&&\kappa_1=2V_{ringo}'(\phi)\left[1+\frac{1}{2}\ln(|\rho|)\right]-\frac{3}{8}V_{paul}(\phi)\rho \nonumber \\
&&\kappa_3=V_{ringo}''(\phi)\ln(|\rho|)-\frac{1}{8}V_{paul}'(\phi)\rho-\frac{1}{4}V_{john}(\phi)\left[1-\ln(|\rho|)\right] \nonumber\\
&&\kappa_8=\frac{1}{2}V_{john}'(\phi)\ln(|\rho|),\qquad  \kappa_9=-\rho_\Lambda^\textrm{bare}-3V_{george}''(\phi)\rho \nonumber
\\
&&F =\frac{1}{2}V_{george}(\phi)-\frac{1}{4} V_{john}(\phi)\rho\ln(|\rho|) \nonumber
\ea
with $V'_{george} \equiv 0$ allowed, if and only if there exist other non-vanishing potentials. It follows that the self-tuning version of Horndeski's theory must take the form
\begin{multline}
S^\textrm{self-tun}=\int d^4 x \left[{\cal L}_{john}+ {\cal L}_{paul}+{\cal L}_{george}+{\cal L}_{ringo}\right. \\\left.-\sqrt{-g}\rho_{\Lambda}^\textrm{bare} \right]
+S_m[g_{\mu\nu}; \Psi_n] \label{selftun}
\end{multline}
where the base Lagrangians are built from {\it the Fab Four} (\ref{eq:john}) to (\ref{eq:ringo}). Note also  the presence of the bare cosmological constant term $\rho_\Lambda^\textrm{bare}$ which can always be absorbed into a renormalisation of the vacuum energy (contained within $S_m$). This serves as a good consistency check of our derivation. Such a term had to be allowed by the self-tuning theories -- if it had not been there it would have amounted to fine tuning the vacuum energy.
\section{The cosmology of {\it the Fab Four}}
We shall now briefly present the cosmological equations for the general self-tuning theory (\ref{selftun}). To this end, we note that the minisuperspace Lagrangians for {\it the Fab Four}  have the desired structure given by equation (\ref{eq:selftunlag}), and that the Friedmann equations describing this cosmology are
\be
{\cal H}_{john}+{\cal H}_{paul}+{\cal H}_{george}+{\cal H}_{ringo}=-\left[\rho_\Lambda+\rho_\textrm{matter}\right]
\ee
where we have absorbed $\rho_\Lambda^\textrm{bare}$ into the vacuum energy contribution $\rho_\Lambda$, and
\ba
&&{\cal H}_{john}=3V_{john}(\phi)\dot\phi^2\left(3H^2+\frac{\kappa}{a^2}\right)) \nonumber\\
&&{\cal H}_{paul}=-3V_{paul}(\phi)\dot\phi^3H\left(5H^2+3\frac{\kappa}{a^2}\right) \nonumber\\
&&{\cal H}_{george}=-6V_{george}(\phi)\left[\left(H^2+\frac{\kappa}{a^2}\right)+H\dot\phi \frac{V'_{george}}{V_{george}}\right]\qquad \nonumber\\
&&{\cal H}_{ringo}=-24V'_{ringo}(\phi)\dot\phi H\left(H^2+\frac{\kappa}{a^2}\right) \nonumber
\ea
The scalar equations of motion are ${\cal E}_{john}+{\cal E}_{paul}+{\cal E}_{george}+{\cal E}_{ringo}=0$ where 
\ba
&&{\cal E}_{john}= 6{d \over dt}\left[a^3V_{john}(\phi)\dot{\phi}\Delta_2\right]  - 3a^3V_{john}'(\phi)\dot\phi^2\Delta_2
 \nonumber\\
&&{\cal E}_{paul}= -9{d \over dt}\left[a^3V_{paul}(\phi)\dot\phi^2H\Delta_2\right]  +3a^3V_{paul}'(\phi)\dot\phi^3H\Delta_2
 \nonumber\\
&&{\cal E}_{george}= -6{d \over dt}\left[a^3V_{george}'(\phi)\Delta_1\right]  +6a^3V_{george}''(\phi)\dot\phi \Delta_1  \nonumber \\
&&\qquad\qquad\qquad\qquad+6a^3V_{george}'(\phi)\Delta_1^2  \nonumber\\
&&{\cal E}_{ringo} = -24 V'_{ringo}(\phi) {d \over dt}\left[a^3\left(\frac{\kappa}{a^2}\Delta_1 +\frac{1}{3}  \Delta_3 \right) \right]  \nonumber
\ea
We see that on-shell-in-a, $H^2=-\kappa/a^2$,  Ringo's contribution to the Friedmann equation loses its dependence on $\dot \phi$. This explains why Ringo cannot self-tune by itself. We should emphasize that  Ringo does not {\it spoil} self-tuning when John and/or  Paul are also present, even though it {\it does} alter the cosmological dynamics. Note also that  if $V'_{george}=0$, and all the other potentials are vanishing, then the scalar equation of motion becomes trivial and does not force self-tuning. 

For a generic combination of {\it the Fab Four} that includes John and/or Paul, we have a scalar tensor model of self-tuning. The self-tuning is forced by the scalar equation of motion, while the gravity equation links phase transitions in vacuum energy to discontinuities in the temporal derivative of the scalar field. On self-tuning vacua, the scalar field is explicitly time dependent, as it must be in order to evade Weinberg's theorem \cite{nogo}. A detailed study of {\it Fab Four} cosmology will be presented elsewhere.
\section{Discussion}
In this letter we have resurrected Horndeski's theory that describes the most general scalar-tensor theory with second order field equations. We have asked which corners of this theory admit a consistent self-tuning mechanism for solving the (old) cosmological constant problem. Remarkably, this reduces the theory down to a combination of four base Lagrangians, dubbed {\it the Fab Four}. Self-tuning is made possible by breaking Poincar\'e invariance in the scalar sector. 

There are hints at some deep underlying structure in this theory. This merits further investigation, but for now we note that each of  {\it the Fab Four} can be  associated with a dimensionally enhanced Euler density. This is immediately evident for George and Ringo, whereas for John and Paul we note that they can both be written in the form $ V(\phi)  \nabla_\mu\phi \nabla_\nu \phi \frac{\delta W}{\delta g^{\mu\nu}}$, with $W_{john}=\int d^4 x\sqrt{-g} R$ and $W_{paul}= -\frac{1}{4}\int d^4 x \sqrt{-g} \phi \hat G$.


Have we {\it really} solved the cosmological constant problem? We have certainly evaded Weinberg's theorem, but there is plenty more to consider. Does the {\it Fab Four} ultimately give rise to a gravity theory that is phenomenologically consistent, in particular at the level of both cosmology and solar system tests. This is a work in progress, but there are reasons to be guardedly optimistic, especially when one considers the fact that John and Paul contain non-trivial derivative interactions that may give rise to a successful Vainshtein effect.  Relevant work involving three of {\it the Fab Four} was carried out in \cite{Amendola:2008vd}.

We should also ask whether or not the self-tuning property of {\it the Fab Four} is spoilt by radiative corrections. Chances are it probably is spoilt by matter loops, but it is interesting to note that the self-tuning is imposed by the scalar equation of motion, and the scalar does not couple directly to matter. The intriguing geometric properties of  {\it the Fab Four} may also play a role here, but such considerations are beyond the scope of this letter. 

In any event, the ethos behind our approach is not to make any grandiose claims regarding a solution of the cosmological constant problem but to ask what can be achieved in this direction at the level of a scalar tensor theory.  Given that our starting point was the most general scalar tensor theory, we should be in a position to make some reasonably general statements. As we have shown,  Weinberg's theorem alone is not enough to rule out possible self-tuning mechanisms, so even if {\it the Fab Four} are ultimately ruled out by other considerations we should be able to say we have learnt something about the obstacles towards solving the cosmological constant problem and how one might think about extending the scope of Weinberg's theorem.

\begin{acknowledgments} 
EJC and AP acknowledge  financial support from the Royal Society and CC from STR-COSMO, ANR-09-BLAN-0157. AP notes that he was born in the same hospital as John Lennon.
\end{acknowledgments}

\appendix

\end{document}